# Controlling the nanoscale rippling of graphene with SiO$_2$ nanoparticles

Z. Osváth, [*ac] E. Gergely-Fülöp,[a] N. Nagy,[a] A. Deák,[a] P. Nemes-Incze,[ac] X. Jin,[bc] C. Hwang,[bc] and L. P. Biró[ac]

[a] Institute of Technical Physics and Materials Science, MFA, Research Centre for Natural Sciences, HAS, 1525 Budapest, P.O. Box 49, Hungary. E-mail: osvath.zoltan@ttk.mta.hu

[b] Center for Nano-metrology, Division of Industrial Metrology, Korea Research Institute of Standards and Science, Yuseong, Daejeon 305-340, Republic of Korea

[c] Korea-Hungary Joint Laboratory for Nanosciences (KHJLN), P.O. Box 49, 1525 Budapest, Hungary

**Abstract**

The electronic properties of graphene can be significantly influenced by mechanical strain. One practical approach to induce strain in graphene is to transfer this atomically thin membrane onto pre-patterned substrates with specific corrugation. The possibility to use nanoparticles to impart extrinsic rippling to graphene has not been fully explored yet. Here we study the structure and elastic properties of graphene grown by chemical vapour deposition and transferred onto a continuous layer of SiO$_2$ nanoparticles with diameters of around 25 nm, prepared on Si substrate by Langmuir-Blodgett technique. We show that the corrugation of the transferred graphene and thus the membrane strain can be modified by annealing at moderate temperatures. The membrane parts bridging the nanoparticles are suspended and can be reversibly lifted by the attractive forces between an atomic force microscope tip and graphene. This allows the dynamic control of the local morphology of graphene nanomembranes.

**Introduction**

The atomically thin graphene membranes are intrinsically non-flat and have random or quasi-periodic corrugations at the nanometer scale.[1,2] Since this is closely affecting the electronic properties, there is an increasing need for the realization of graphene sheets with controlled corrugation. Substrates play a crucial role, as the graphene–substrate interaction can impart an extrinsic rippling to graphene which differs from its intrinsic corrugation.[3,4] Such rippling can contribute to the scattering of charge carriers.[5,6] In order to preserve the high carrier mobility needed for nanoelectronic applications, atomically flat mica[7] and hexagonal boron nitride[8] substrates were introduced recently, which reduce charge inhomogeneity[9] and smooth out corrugations in graphene leading to ultra-flat morphology. On the other hand, corrugated graphene can be good candidate for sensor applications, as recent simulations[10,11] predict enhanced chemical activity in rippled graphene. The crests and troughs of graphene ripples form active sites for the adsorption or chemisorption of different molecules. It was proposed – based on first-principles calculations[12] – that this can open a way for tunable, regio-selective functionalization of graphene. The extrinsic rippling can be induced for example by pre-prepared elastic substrates[13] or silica nanoparticles (NPs),[14] a possibility which has not been fully explored yet experimentally.[15] In this work we investigate by atomic force microscopy (AFM) the properties of graphene flakes transferred onto a continuous layer of SiO$_2$ NPs, and show that the extrinsic graphene rippling can be controlled by annealing. Due to the high nanoparticle density, graphene membranes remain completely detached from the Si substrate. We were able to map the suspended graphene parts bridging the nanoparticles by carefully adjusting the AFM imaging parameters. Local indentation was performed on the suspended parts in order to investigate the elastic properties of the graphene membrane.

**Experimental**

Amorphous silica NPs were synthesized according to the Stöber-method, which we used earlier to prepare NPs with different diameters.[16,17] In this work, silica nanospheres with ~25 nm diameter were prepared as follows. First, a solution containing 50 ml ethanol (absolute, VWR), 1.594 ml NH3 (32 %; Scharlau) and 0.44625 ml H$_2$O (ultrapure, resistivity: 18.2 MOhm/cm) was stirred for 30 minutes. Then, 2 ml tetraethyl orthosilicate (reagent grade 98 %; Aldrich) was added to this solution and stirred overnight. Finally the ammonia was removed by distillation at 60 °C.

Langmuir-Blodgett (LB) films of the nanoparticles were prepared in a KSV 2000 film balance. The ethanolic solution of NPs was sonicated for 5 minutes, then mixed with chloroform (Scharlau, reagent grade, stabilized with ethanol) and spread at the air/water interface. After 30 minutes the particles were compressed at a barrier speed of 0.4 cm$^2$/s. After the surface pressure reached ~1 mN/m, the speed was lowered to 0.2 cm$^2$/s. The LB films were prepared by vertical deposition (6 mm/min) at ca. 80 % of the collapse pressure, which was measured before. We used silicon slices as substrates, which were cleaned with acetone, water, 2 % hydrofluoric acid solution, and finally rinsed in water.

Graphene was grown on a mechanically and electro-polished copper foil (25 μm thick, 99.8% purity, Alfa-Aesar) which was inserted into a thermal CVD furnace. The furnace was evacuated to ~10$^{-4}$ torr and the temperature was raised to 1010 °C with H$_2$ gas flow (~10$^{-2}$ torr). When the temperature became stable, both CH$_4$ (20 sccm) and H$_2$ (5 sccm) were injected into the furnace for 8 minutes to synthesize the graphene. After the growth, we cooled down the furnace with a cooling rate of 50 °C/min.

The graphene sample was transferred onto the SiO$_2$ NPs using thermal release tape, and an etchant mixture consisting of CuCl$_2$ aqueous solution (20%) and hydrochloric acid (37%) in 4:1 volume ratio. After the etching procedure, the tape holding the graphene was rinsed in distilled water, then dried and pressed onto the surface covered by nanoparticles. The tape/graphene/SiO$_2$NPs/Si sample stack was placed on a hot plate and heated to 5 °C above the 90 °C release temperature of the tape. The tape was removed, leaving behind the graphene on top of SiO$_2$ NPs. This was confirmed by confocal Raman microscopy using an excitation laser of 488 nm. The sample was annealed at 400 °C in N$_2$ atmosphere for 2 hours in order to improve the adhesion of graphene to the NPs.

The sample was investigated both before and after annealing by confocal Raman microscopy and a MultiMode 8 AFM from Bruker operating under ambient conditions. Both conventional Tapping and Peak Force Tapping modes were used. Sharp silicon cantilevers were applied with tip radius R ≃ 2 nm and spring constant k = 9.2 N/m. Peak Force Tapping is a relatively new scanning mode available with the MultiMode 8 AFM, where a complete force-distance curve is performed in every measuring point, while the z-piezo data of the cantilever is recorded at the maximal force between the sample and the cantilever. This maximal force defines a setpoint for image acquisition and can be changed in order to record images at different sample-cantilever forces. To investigate the mechanical properties of the CVD-grown graphene sample, we used a stiffer AFM cantilever with tip radius R ≃ 8 nm and spring constant k = 34 N/m, as determined in situ by the thermal tune method,[18] prior to indentation experiments.

**Results and discussion**

Graphene was successfully transferred on top of $SiO_2$ NPs as seen in Fig. 1a, which shows the confocal Raman map of the graphene *2D* peak intensity. Note that the graphene is not continuous. It is split (along the dark stripes) into sheets with different sizes, typically of several micrometers. This splitting is attributed to the dry transfer procedure using thermal release tape. Figure 1b shows the average Raman spectrum of the graphene sheets mapped in Fig. 1a. The typical graphene peaks (*D*, *G*, and *2D*) are labelled in the spectrum.

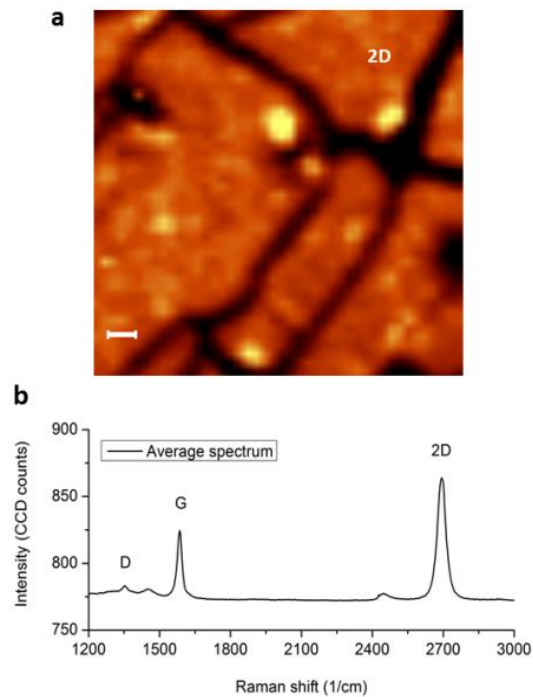

**Fig. 1** Confocal Raman microscopy of transferred graphene. (a) Raman map of the 2D graphene peak intensity. Scalebar is 500 nm. The dark lines correspond to the substrate not covered with graphene. (b) Average spectrum of the graphene sheets shown in (a).

In the following we analyse in more details the *2D* peak. When graphene is transferred onto conventional $SiO_2$/Si substrate (Fig. 2a-b), the *2D* peak measured under the laser spot (Fig. 2a) is very well fitted with a Lorentzian function, which gives a full width at half maximum (FWHM) of $w_L = 25.8\ cm^{-1}$. If we now consider the average of many Raman spectra measured on a larger area (5×5 μm²) of the $SiO_2$/Si substrate, the average *2D* peak (Fig. 2b) is broadened due to inhomogeneous distribution in the sample (local strain and doping effects[19]).

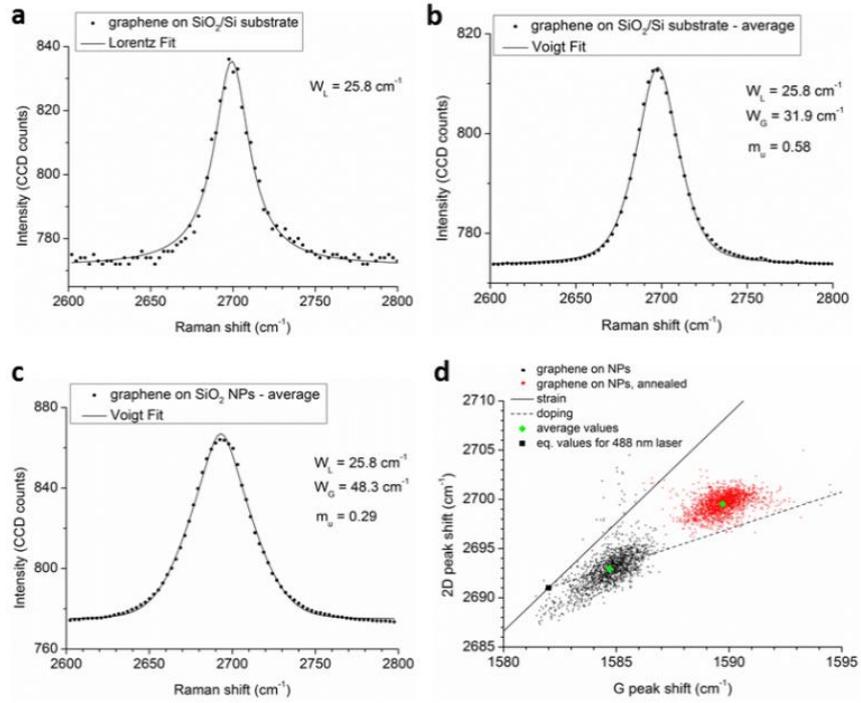

**Fig. 2** Raman spectra of transferred graphene. (a) Lorentzian fit to a *2D* peak of graphene on SiO$_2$/Si substrate measured in one point. (b) Voigt fit to the *2D* graphene peak averaged on a 5×5 μm$^2$ area on SiO$_2$/Si substrate. (c) Voigt fit to the *2D* graphene peak averaged on a 5×5 μm$^2$ area on SiO$_2$ nanoparticles (Fig. 1). (d) ($\omega_G, \omega_{2D}$) correlation plot before (black dots) and after annealing (red dots). The corresponding average peak positions are marked with green diamonds. The equilibrium values for 488 nm laser are shown with a black square. The slopes denoting purely strain (straight line) and purely doping effects (dashed line) are also plotted.

This average spectrum is fitted with a pseudo-Voigt peak function, which is a linear combination of a Gaussian and Lorentzian function and describes the Gaussian broadening of a Lorentz peak characterized with w$_L$:

$$y = y_0 + A \left[ m_u \frac{2}{\pi} \frac{w_L}{4(x-x_c)^2 + w_L^2} + (1-m_u) \frac{\sqrt{4\ln 2}}{\sqrt{\pi}\, w_G} e^{-\frac{4\ln 2}{w_G^2}(x-x_c)^2} \right]$$

Here *A* is the peak amplitude, $x_c$ the peak centre, w$_G$ is the Gaussian FWHM, and $m_u$ is the profile shape factor. In the case of a pure Lorentzian line shape $m_u = 1$. The averaged peak is broadened due to small variations in the spectral position of individual peaks. This is taken into account as a Gaussian distribution. For graphene on SiO$_2$/Si substrate, w$_G = 31.9\ cm^{-1}$, and $m_u = 0.58$, which shows that the Lorentzian component is still more important than the Gaussian one $(1 - m_u)$. This is not the case when we transfer the graphene onto SiO$_2$ nanoparticles. Fig. 2c shows the Raman *2D* peak of graphene transferred onto SiO$_2$ NPs, and averaged on an area of 5×5 μm$^2$ (Fig. 1). The peak is very well fitted with the pseudo-Voigt function which yields w$_G = 48.3\ cm^{-1}$, reflecting a more significant broadening of the Lorentz peak (w$_L = 25.8\ cm^{-1}$). The profile shape factor is $m_u = 0.29$, half the value obtained on conventional SiO$_2$/Si substrate. This shows that in this case the Gaussian component is much more important. In order to study the origin of this Gaussian distribution, we constructed the correlation plot ($\omega_G, \omega_{2D}$) from the *G* peak positions ($\omega_G$) and the *2D* peak positions measured on the area in Fig. 1. This correlation plot is shown by black dots in Fig. 2d. Additionally, we plotted the slopes $\Delta\omega_{2D}/\Delta\omega_G$ corresponding to variations induced by strain only (solid line) and by purely doping effects (dotted line), respectively. We used ($\Delta\omega_{2D}/\Delta\omega_G$) = 2.2 for the strain slope, and ($\Delta\omega_{2D}/\Delta\omega_G$) = 0.75 for the p-type doping slope.[19] One can observe that the peak positions are shifted from the equilibrium values $(\omega_G^0, \omega_{2D}^0) = (1582\ cm^{-1}, 2691\ cm^{-1})$[20] denoted by black square in Fig. 2d. The average peak positions obtained from the black dots are $(1584.7\ cm^{-1}, 2692.9\ cm^{-1})$, which is a point located very close to the doping slope (lower green diamond symbol) on the correlation plot. This shows that spatial doping inhomogeneity plays important role in the peak shifts observed on graphene transferred onto NPs. After annealing the sample at 400 °C, we performed the same study by confocal Raman microscopy. The red dots in Fig. 2d are extracted from spectra acquired from an area of 5×5 µm². One can immediately see the large peak shifts towards higher wavenumbers. The average positions are 1589.7 $cm^{-1}$ and 2699.5 $cm^{-1}$ for the *G* and the *2D* peaks, respectively, which is denoted by the corresponding green diamond symbol on the correlation plot. Note, that this point is now located farther from the doping slope, indicating that annealing introduced some strain in the graphene membrane. We can estimate the strain ($\Delta\varepsilon$) using $\Delta\varepsilon = -\Delta\omega_{2D}/(2\omega_{2D}^0\gamma_{2D})$[21], where $\gamma_{2D} = 2.7$ is the Grüneisen parameter of the *2D* peak obtained from first-principles calculations.[22] Using $\Delta\omega_{2D} = 6.6\ cm^{-1}$, the difference between average *2D* peak positions obtained before and after annealing, and neglecting the contribution from doping, we obtain an average compressive strain of $\Delta\varepsilon = -0.045\%$. In order to see the effect of higher temperatures, we further annealed the sample at 550 °C for two hours. Confocal Raman measurements performed after the second annealing show that the above average strain could not be increased significantly.

In order to investigate the microscopic details of this strained graphene membrane, we performed tapping mode AFM in the following cases: (a) – as prepared LB film of $SiO_2$ NPs (Fig. 3a); (b) – $SiO_2$ NPs covered with graphene (Fig. 3b); (c) – $SiO_2$ NPs annealed at 400 °C, without graphene (Fig. 3c); and (d) – $SiO_2$ NPs covered with graphene and annealed at 400 °C (Fig. 3d). Figure 3a shows that the NPs cover completely the Si substrate. The resulting surface can be characterized with an RMS value of 2.94 nm. After transferring graphene on the top of NPs the surface RMS value is slightly reduced (2.26 nm, Fig. 3b) and the shape of the NPs is not clearly resolved in the AFM image. This is because graphene is loosely bound to the NPs and does not follow closely the surface morphology. In order to promote the adhesion[23] to NPs, we annealed the sample at 400 °C in $N_2$ atmosphere. We observed a small rearrangement of the NPs after annealing, due to which the uncovered, nanoparticle-free areas of the Si substrate slightly increased (dark regions in Fig. 3c). These uncovered areas allowed for the measurement of nanoparticle diameters. For example, the line section labelled *1* (Fig. 3c and Fig. 3e) reads a height difference of 25.9 nm between the vertical dashed lines, which approximately corresponds to the diameter of the measured NPs. Note that the RMS of the surface increased to about 4.73 nm due to the apparition of NP-free Si areas. Furthermore, the RMS of the graphene-covered regions is around 3.27 nm after annealing, which is 70% larger than the value measured before annealing (2.26 nm). As the AFM image in Fig. 3d shows, this is attributed to the fact that upon annealing the graphene morphology adapts to take the shape of the NPs.[23] As a result the graphene membrane conforms better to the nanoparticle-induced surface corrugation, and this induces the compressive strain determined by confocal Raman microscopy. It is important to note that in this case the graphene bridges the NP-free areas, and significant suspended graphene areas are produced. The line section labelled *1'* in Fig. 3d and Fig. 3f reads a height difference of only 9.5 nm between the vertical lines, clearly showing that graphene do not reach the NP-free Si substrate, but instead is suspended between the neighbouring NPs, forming a graphene hammock.

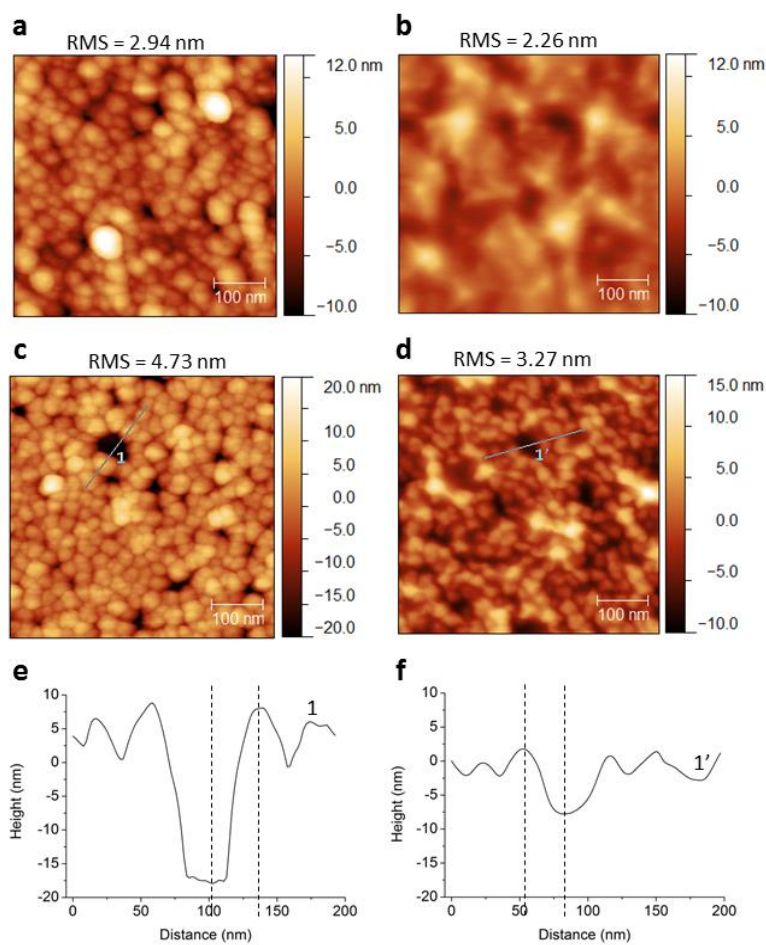

**Fig. 3** Tapping mode AFM images of SiO$_2$ NPs. (a) As prepared by LB technique. (b) Covered with graphene. (c) Annealed at 400 °C, without graphene. (d) Covered with graphene and annealed at 400 °C. (e) The line section labelled 1 in (c). The vertical distance between the substrate and the top of NPs (dashed lines) is 25.9 nm. (f) The line section labelled 1' in (d). The vertical distance between the dashed lines is 9.5 nm, showing a graphene membrane suspended between NPs.

It is worth noticing that the NPs covered with graphene give lower phase signal in the tapping mode AFM investigations, compared to the bare NP surface, thus the phase images can be used to unambiguously identify graphene-covered regions in large area scans (see Fig. 4).

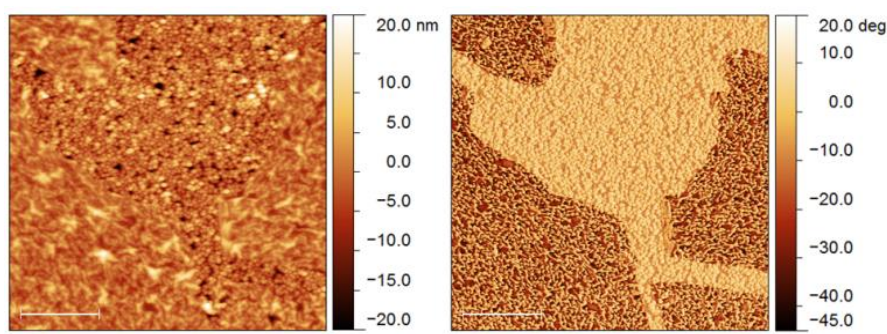

**Fig. 4** Tapping mode AFM images of SiO$_2$ NPs partially covered with graphene (annealed sample). Scale bars are 500 nm. The left image is the topography, while the right image is the AFM phase map from the same area. Graphene-covered regions (dark coloured phase) can be unambiguously distinguished from bare nanoparticles (light coloured phase).

Furthermore, we observed that by changing the scanning parameters we can reveal important details in both the phase and the topographic AFM images of the annealed sample. For example, in Fig. 5a we show the AFM image of an area of 400×400 nm$^2$ with graphene-covered NPs, acquired with 62 mV drive amplitude and setpoint of 350 mV. The free amplitude of the cantilever was 500 mV. Several low-phase (dark) spots appear on the phase image (right panel), which apparently are randomly distributed. In parallel, height jumps appear in the same spots on the topographic image (left panel). By increasing the setpoint to 425 mV (Fig. 5b), extended low-phase areas appear on the phase image of the same 400×400 nm$^2$ area (right panel), while higher *z*-values (height jumps) are also measured on these areas on the topographic image (left panel). We illustrate this effect quantitatively by plotting in Fig. 6 the height profiles of the chosen line sections labelled *1–1'* (Fig. 5a-b), corresponding to amplitude setpoints of 350 mV and 425 mV, respectively. The height profiles reveal a vertical difference of about 2 nm between sections *1* and *1'*.

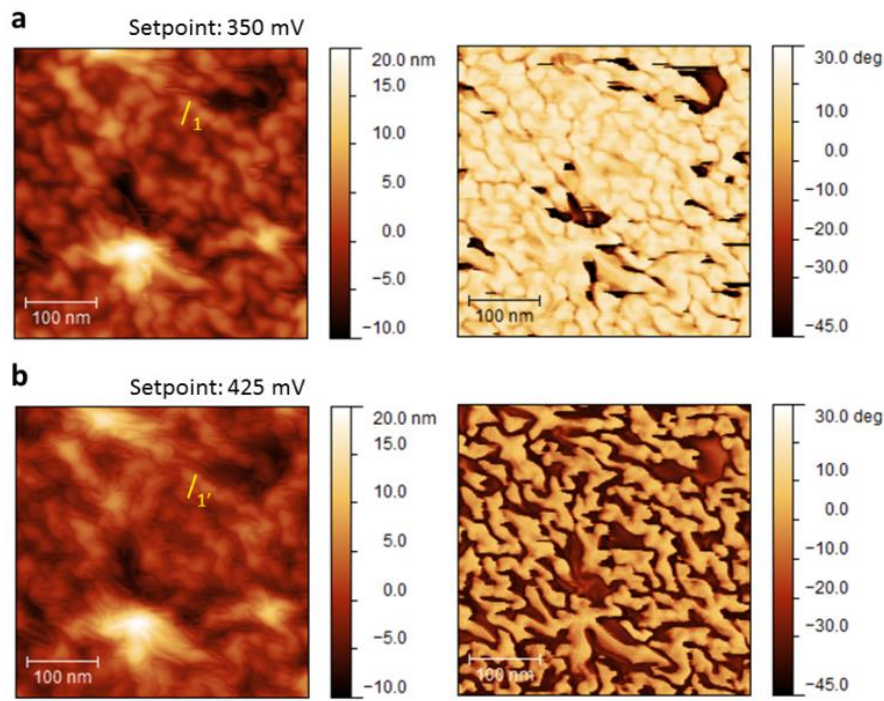

**Fig. 5** Tapping mode AFM images of graphene on top of SiO$_2$ NPs (annealed sample). Topographic images are shown on the left, while the corresponding phase images are displayed on the right. The same area was measured with amplitude setpoints of (a) 350 mV, and (b) 425 mV. Low-phase areas reveal suspended graphene parts. The line sections 1 – 1' are shown in Fig. 6.

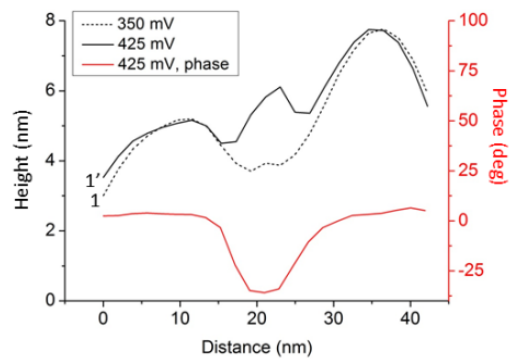

**Fig. 6** Topographic height profiles along the line sections labelled 1 (dashed line) and 1' (black line) from Fig. 5, which show the same graphene part measured with AFM amplitude setpoints of 350 mV and 425 mV, respectively. Additionally, the phase signal corresponding to profile 1' is also displayed (red line), showing decreased phase values at the suspended graphene regions.

The plot in Fig. 6 contains also the phase signal corresponding to the line section *1'*. Note that the phase signal is decreased at the place where the height jump occurs. This decreased phase shows a modified interaction between graphene and the AFM cantilever. Comparing the topography and phase maps we identify the low-phase areas as the graphene regions suspended between $SiO_2$ nanoparticles. By increasing the setpoint to 425 mV, we actually lowered the interaction force between graphene and cantilever. As a result, at this setting the van der Waals attractive force became dominant and pulled up the suspended graphene parts, when scanning over them, producing height jumps of about 2 nm in the topographic images. This resulted also in a modified phase signal. The effect is similar to the bistable and oscillatory motion of graphene nanomembrane observed by scanning tunneling microscopy (STM) at the scale of an intrinsic rippling (~3 nm).[24] STM tip-induced deformation of graphene was observed also at larger scales.[25,26] Recent experiments show that the extrinsic rippling of graphene can also be enhanced by the electric field of an STM tip.[27] In our case, we were able to reveal by AFM extended graphene regions suspended between silica nanoparticles. It is worthy to note that the lifting of suspended graphene parts can be completely avoided by increasing the drive amplitude to 80 – 90 mV, as it can be seen in Fig. 3b and Fig. 3d.

Next, we investigated the elastic properties of the CVD-grown graphene sample. Nanoindentation of graphene was performed on suspended areas of around 70 – 100 nm in diameter. One of these areas is shown in Fig. 7a, which is similar to the ones discussed previously in Fig. 3d.

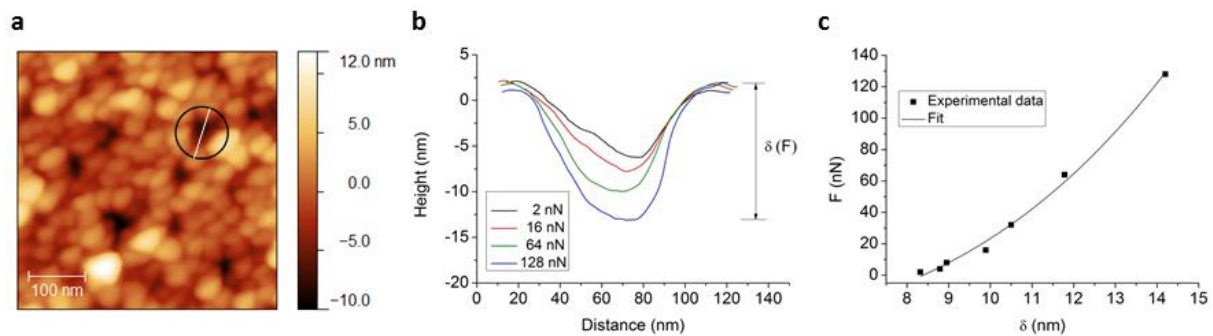

**Fig. 7** Nanoindentation performed in Peak Force AFM mode. (a) Topographic image of graphene-covered $SiO_2$ NPs acquired at a peak force of *F* = 16 nN. (b) Height profiles taken along the same line section (white line) in (a), measured at different load forces (*F*). *δ* is the force-induced deflection. (c) Force-deflection data.

Nanoindentation experiments were performed in Peak Force mode with an AFM cantilever with tip radius $R \simeq 8$ nm and spring constant k = 34 N/m. The same area was scanned repeatedly by increasing gradually the Peak Force setpoint from 2 nN to 128 nN. A complete image was recorded for every force setpoint (*F*). Selected height profiles are shown in Fig. 7b, which were extracted from the images recorded at the corresponding tip-sample force values. All profiles were taken along the same line section shown in Fig. 7a (white line), which shows the AFM image acquired at *F* = 16 nN. The force-induced deflection (*δ*) of the suspended graphene nanomembrane was measured as the difference between crests and troughs of the height profiles. Force-deflection data are obtained, as shown in Fig. 7c. Note that the *δ* = 8.3 nm measured at *F* = 2 nN is the initial deflection of the graphene hammock and is considered as an offset in further analysis. It is also worth noticing that the deflection induced by larger forces is reversible and the indentation does not lead to permanent deformation of the graphene membrane. To interpret the experimental data, we used the indentation model of a circular monolayer graphene by a spherical indenter.[28] The graphene area considered is marked by a circle in Fig. 7a, which has a radius of approximately $a$ = 50 nm. The nominal radius of the AFM tip is $R \simeq 8$ nm. We fitted the data with $F = c\delta + d\delta^3$,[28–31] where the coefficients $c$ and $d$ are related to the Young's modulus $E$ and pre-tension $\sigma_0$ of a membrane of thickness $h$ (0.34 nm for graphene):

$$c = \sigma_0 \pi h, \quad d = Eq^3 a^{-2} h (R/a)^{1/4}. \qquad (1)$$

Here $q = 1/(1.05 - 0.15\nu - 0.16\nu^2) = 0.98$, with $\nu = 0.165$ the Poisson's ratio for monolayer graphene.[29,32] In our case $R/a = 0.16$ (see also Table 1, area no. 3), and we already took into account a correction factor of $(R/a)^{1/4}$ in Eq. (1), proposed if $R/a > 0.14$ (sphere load model).[33,34] The fit gives coefficient values ($c$, $d$) = (0.1115, 0.0564) from which we obtain $E = 0.69$ TPa, and a pre-tension of $\sigma_0 = 0.1$ GPa. We performed the same measurements on several – similarly suspended – graphene areas, and calculated the Young's modulus as above. The results are shown in Table 1.

| Area no. | R/a | d | E (TPa) |
|---|---|---|---|
| 1 | 0.141 | 0.0298 | 0.48 |
| 2 | 0.158 | 0.0833 | 1.05 |
| 3 | 0.16 | 0.0564 | 0.69 |
| 4 | 0.163 | 0.0363 | 0.43 |
| 5 | 0.174 | 0.1142 | 1.17 |
| 6 | 0.183 | 0.0965 | 0.87 |
| 7 | 0.184 | 0.0748 | 0.67 |
| 8 | 0.186 | 0.1399 | 1.22 |
| 9 | 0.19 | 0.0719 | 0.6 |
| 10 | 0.194 | 0.1085 | 0.86 |
| 11 | 0.202 | 0.0919 | 0.67 |
| 12 | 0.207 | 0.0775 | 0.53 |
| 13 | 0.208 | 0.2054 | 1.4 |
| 14 | 0.232 | 0.2364 | 1.25 |
| 15 | 0.24 | 0.2572 | 1.27 |

**Table 1** Nanoindentation experiments performed on suspended graphene areas with diameter $2a$. The Young's modulus ($E$) is calculated using the fitting parameter $d$. The tip radius is $R$ ≈ 8 nm.

Note that in all cases $R/a > 0.14$, which satisfies the requirement of the sphere load model. Calculating the average of the Young's moduli shown in Table 1 we obtain $E_{avg}$ = 0.88 TPa. This value is 12% smaller than the expected value of 1 TPa determined by recent experiments on both CVD-grown[35] and exfoliated samples.[29,36,37] The reason for that we think is related to the fact that no appropriate deflection data could be measured in the high load regime. At load forces higher than 200 nN the deflection values are comparable to the NP's diameter ($\delta \simeq 25$ nm), i.e. the graphene reaches the Si substrate. In this force range the tip apex of the AFM cantilever starts blunting ($R$ increases) which also affects the measurements. Nevertheless, and even though deflection data are not available in the high load regime, in some of the cases (area no. 2, 5, 8, 13, 14, and 15 in Table 1) we obtained $E$ values close to or even higher than 1 TPa. These results show that the elastic properties of graphene can be very well studied by Peak Force AFM measurements on suspended nanomembranes of 70 – 100 nm in diameter ($a$ = 35 – 50 nm), which is one order of magnitude less than in previous experiments performed on graphene membranes.[29,35]

**Conclusions**

In summary, we have investigated by AFM and confocal Raman microscopy the properties of CVD-grown graphene transferred onto a Langmuir-Blodgett film of $SiO_2$ nanoparticles. We showed that the nanoscale rippling of graphene can be modified by annealing at moderate temperatures (400 $^o$C), which introduces compressive strain into the atomically thin membrane. Both topographic and phase images revealed extended graphene regions suspended between silica nanoparticles. This gave the possibility to investigate by local indentation the elastic properties of the transferred graphene. Regulating the extrinsic morphology of graphene by nanoparticles opens new pathways to fine tune the propeties of graphene. These may include regioselective functionalization[10,12] or tunable molecular doping.[11] Here we presented a method for the preparation and mapping of suspended graphene regions. The dynamic control of the local graphene morphology can play an important role in the development of graphene based nanomechanical devices such as switches.[38–40]

**Acknowledgements**

The research leading to these results has received funding from the Korean-Hungarian Joint Laboratory for Nanosciences and the People Programme (Marie Curie Actions) of the European Union's Seventh Framework Programme under REA grant agreement n° 334377. The OTKA grants K101599, PD-105173 in Hungary, as well as the János Bolyai Research Fellowships from the Hungarian Academy of Sciences are acknowledged.

**References**

1  A. Fasolino, J. H. Los and M. I. Katsnelson, *Nat. Mater.*, 2007, **6**, 858.
2  J. C. Meyer, A. K. Geim, M. I. Katsnelson, K. S. Novoselov, T. J. Booth and S. Roth, *Nature*, 2007, **446**, 60.
3  M. Ishigami, J. H. Chen, W. G. Cullen, M. S. Fuhrer and E. D. Williams, *Nano Lett.*, 2007, **7**, 1643.
4  V. Geringer, M. Liebmann, T. Echtermeyer, S. Runte, M. Schmidt, R. Rückamp, M. C. Lemme and M. Morgenstern, *Phys. Rev. Lett.*, 2009, **102**, 076102.
5  M. I. Katsnelson and A. K. Geim, *Phil. Trans. R. Soc. A*, 2008, **366**, 195.
6  G.-X. Ni, Y. Zheng, S. Bae, H. R. Kim, A. Pachoud, Y. S. Kim, C.-L. Tan, D. Im, J.-H. Ahn, B. H. Hong and B. Özyilmaz, *ACS Nano*, 2012, **6**, 1158.
7  C. Lui, L. Liu, K. Mak, G. Flynn and T. Heinz, *Nature*, 2009, **462**, 339.
8  C. R. Dean, A. F. Young, I. Meric, C. Lee, L. Wang, S. Sorgenfrei, K. Watanabe, T. Taniguchi, P. Kim, K. L. Shepard and J. Hone, *Nat. Nanotechnol.*, 2010, **5**, 722.
9  R. Decker, Y. Wang, V. W. Brar, W. Regan, H.-Z. Tsai, Q. Wu, W. Gannett, A. Zettl and M. F. Crommie, *Nano Lett.*, 2011, **11**, 2291.
10 D. W. Boukhvalov and M. I. Katsnelson, *J. Phys. Chem. C*, 2009, **113**, 14176.
11 D. W. Boukhvalov, *Surf. Sci.*, 2010, **604**, 2190.
12 X. Gao, Y. Wang, X. Liu, T.-L. Chan, S. Irle, Y. Zhao and S. B. Zhang, *Phys. Chem. Chem. Phys.*, 2011, **13**, 19449.
13 S. Scharfenberg, D. Z. Rocklin, C. Chialvo, R. L. Weaver, P. M. Goldbart and N. Mason, *Appl. Phys. Lett.*, 2011, **98**, 091908.
14 M. Yamamoto, O. Pierre-Louis, J. Huang, M. S. Fuhrer, T. L. Einstein and W. G. Cullen, *Phys. Rev. X.*, 2012, **2**, 041018.
15 T. Li, *Modelling Simul. Mater. Sci. Eng.*, 2011, **19**, 054005.
16 A. Deák, E. Hild, A. L. Kovács and Z. Hórvölgyi, *Phys. Chem. Chem. Phys.*, 2007, **9**, 6359.
17 A. Deák, B. Bancsi, A. L. Tóth, A. L. Kovács and Z. Hórvölgyi, *Colloid. Surface. A.*, 2006, **278**, 10.

18 J. L. Hutter and J. Bechhoefer, *Rev. Sci. Instrum.*, 1993, **64**, 1868.
19 J. E. Lee, G. Ahn, J. Shim, Y. S. Lee and S. Ryu, *Nat. Commun.*, 2012, **3**, 1024.
20 I. Calizo, A. A. Balandin, W. Bao, F. Miao and C. N. Lau, *Nano Lett.*, 2007, **7**, 2645.
21 F. Ding, H. Ji, Y. Chen, A. Herklotz, K. Dörr, Y. Mei, A. Rastelli and O. G. Schmidt, *Nano Lett.*, 2010, **10**, 3453.
22 T. M. G. Mohiuddin, A. Lombardo, R. R. Nair, A. Bonetti, G. Savini, R. Jalil, N. Bonini, D. M. Basko, C. Galiotis, N. Marzari, K. S. Novoselov, A. K. Geim and A. C. Ferrari, Phys. Rev. B, 2009, **79**, 205433.
23 S. Pang, J. M. Englert, H. N. Tsao, Y. Hernandez, A. Hirsch, X. Feng and K. Müllen, *Adv. Mater.*, 2010, **22**, 5374.
24 T. Mashoff, M. Pratzer, V. Geringer, T. J. Echtermeyer, M. C. Lemme, M. Liebmann and M. Morgenstern, *Nano Lett.*, 2010, **10**, 461.
25 N. N. Klimov, S. Jung, S. Zhu, T. Li, C. A. Wright, S. D. Solares, D. B. Newell, N. B. Zhitenev and J. A. Stroscio, *Science*, 2012, **336**, 1557.
26 P. Xu, Y. Yang, S. D. Barber, M. L. Ackerman, J. K. Schoelz, D. Qi, I. A. Kornev, L. Dong, L. Bellaiche, S. Barraza-Lopez and P. M. Thibado, *Phys. Rev. B*, 2012, **85**, 121406.
27 Z. Osváth, F. Lefloch, V. Bouchiat and C. Chapelier, *Nanoscale*, 2013, **5**, 10996.
28 X. Tan, J. Wu, K. Zhang, X. Peng, L. Sun and J. Zhong, *Appl. Phys. Lett.*, 2013, **102**, 071908.
29 C. Lee, X. Wei, J. W. Kysar and J. Hone, *Science*, 2008, **321**, 385.
30 K. T. Wan, S. Guo and D. A. Dillard, *Thin Solid Films*, 2003, **425**, 150.
31 U. Komaragiri, M. R. Begley and J. G. Simmonds, *J. Appl. Mech.*, 2005, **72**, 203.
32 O. L. Blakeslee, D. G. Proctor, E. J. Seldin, G. B. Spence and T. Weng, *J. Appl. Phys.*, 1970, **41**, 3373.
33 M. R. Begley and T. J. Mackin, *J. Mech. Phys. Solid.*, 2004, **52**, 2005.
34 O. N. Scott, M. R. Begley, U. Komaragiri and T. J. Mackin, *Acta Mater.*, 2004, **52**, 4877.
35 G.-H. Lee, R. C. Cooper, S. J. An, S. Lee, A. van der Zande, N. Petrone, A. G. Hammerberg, C. Lee, B. Crawford, W. Oliver, J. W. Kysar and J. Hone, *Science*, 2013, **340**, 1073.
36 S. P. Koenig, N. G. Bodetti, M. L. Dunn and J. S. Bunch, *Nat. Nanotechnol.*, 2011, **6**, 543.
37 J. S. Bunch, S. S. Verbridge, J. S. Alden, A. M. van der Zande, J. M. Parpia, H. G. Craighead and P. L. McEuen, *Nano Lett.*, 2008, **8**, 2458.
38 J. S. Bunch, A. M. van der Zande, S. S. Verbridge, I. W. Frank, D. M. Tanenbaum, J. M. Parpia, H. G. Craighead and P. L. McEuen, *Science*, 2007, **315**, 490.
39 K. M. Milaninia, M. A. Baldo and A. Reina, J. Kong, *Appl. Phys. Lett.*, 2009, **95**, 183105.
40 P. Li, Z. You, G. Haugstad and T. Cui, *Appl. Phys. Lett.*, 2011, **98**, 253105.